\documentclass[prd, 10pt,aps,nofootinbib, floatfix, notitlepage,twocolumn]{revtex4-1}
\usepackage{amsmath,amsfonts,amssymb}
\usepackage{mathrsfs}
\usepackage{multirow}
\usepackage{graphicx}
\usepackage[english]{babel} 
\usepackage{url} 
\usepackage{color}
\usepackage[utf8]{inputenc} 
\usepackage{float}
\usepackage[colorlinks=true,citecolor=blue,urlcolor=blue]{hyperref}
\usepackage{booktabs}
\usepackage{subfigure}

\usepackage[T1]{fontenc}

\begin{document}


\title{Redshift-dependent Distance Duality Violation in Resolving Multidimensional Cosmic Tensions}


\author{Zhihuan Zhou}
\email{zhihuanzhou1@163.com}
\author{Zhuang Miao}%
\author{Rong Zhang}%
\author{Hanbing Yang}%
\author{Penghao Fu}%
\author{Chaoqian Ai}%
\email{acq0003@live.com}
\affiliation{
	School of Engineering \\
	Xi'an International University \\
	Xi'an 710077, People's Republic of China}%


\date{\today}

\begin{abstract}
	
	In this work, we investigate whether violations of the distance-duality relation (DDR) can resolve the multidimensional cosmic tensions characterized by the $H_0$ and $S_8$ discrepancies. Using the Fisher-bias formalism, we reconstruct minimal, data-driven $\eta(z)$ profiles that capture the late-time deviations required to reconcile early- and late-Universe calibrations. While a constant DDR offset preserves the Pantheon-inferred matter density $\Omega_m = 0.334 \pm 0.018$—leaving its inconsistency with the \emph{Planck} best-fit $\Lambda$CDM model and weak-lensing surveys unresolved—a time-varying DDR substantially reduces cross-dataset inconsistencies and improves the global fit, yielding $\Delta\chi^2 \simeq -10$ relative to $\Lambda$CDM when the SH0ES prior is excluded. This result suggests that the $\Omega_m$ discrepancy may represent indirect evidence for a time-varying DDR. A hybrid scenario combining a time-dependent DDR with a phantom-like dark energy transition achieves the most consistent global reconciliation, reducing the tension with DES-Y3 measurements to below $2\sigma$. These findings indicate that a mild DDR violation, coupled with evolving dark energy, offers a coherent pathway toward jointly addressing the $H_0$ and $S_8$ tensions.
\end{abstract}


\maketitle

\section{Introduction}

The $\Lambda$ Cold Dark Matter ($\Lambda$CDM) paradigm has proven remarkably successful in describing the expansion history and structure formation of the Universe, supported by high-precision fits to diverse cosmological observations~\cite{SupernovaSearchTeam:1998fmf,Planck:2018vyg,Planck:2019nip,Abbott:2017wau,Alam:2016hwk,DESI:2025zgx,DES:2024jxu}. Nevertheless, its two primary components—dark energy and dark matter—remain without direct physical explanation. Moreover, persistent discrepancies among datasets have emerged, challenging the internal consistency of the $\Lambda$CDM framework~\cite{CosmoVerse:2025txj,Poulin:2023lkg,Verde:2023lmm,Verde:2019ivm,DiValentino:2021izs}. Most notably, the Hubble tension—a $>5\sigma$ discrepancy between early-Universe estimates of $H_0$ from the \emph{Planck} cosmic microwave background (CMB) ($67.36 \pm 0.54$ km/s/Mpc)~\cite{Aghanim:2018eyx} and late-Universe measurements calibrated via Cepheids ($73.17 \pm 0.86$ km/s/Mpc) by SH0ES~\cite{Riess:2024vfa}—remains a compelling signal of potential new physics. Alongside it, the $S_8$ tension, characterized by systematically lower clustering amplitudes in weak-lensing surveys compared to CMB-based predictions, further motivates a re-examination of $\Lambda$CDM assumptions regarding cosmic distances and growth~\cite{DES:2021bvc,KiDS:2020ghu,Kilo-DegreeSurvey:2023gfr,Sabogal:2024yha,Raveri:2023zmr,DiValentino:2025sru}.

A growing body of work \cite{Camarena:2023rsd,Poulin:2024ken,Pedrotti:2024kpn,Teixeira:2025czm} interprets the Hubble tension as a calibration mismatch between Type~Ia supernovae (SNe~Ia) and baryon acoustic oscillations (BAO) when both are anchored to absolute distance scales—specifically, the Cepheid-based luminosity calibration ($M_B$)~\cite{Alestas:2020zol,Marra:2021fvf} and the CMB-inferred sound horizon ($r_d$). This \textit{cosmic calibration tension} arises when angular diameter distances $D_A(z)$ inferred from BAO are converted into luminosity distances $D_L(z)$ via the distance-duality relation (DDR) and subsequently compared with the SNe~Ia distance scale. In practice, the tension originates from two key anchors: (i) the \emph{Pantheon+} SNe~Ia catalogue calibrated using the SH0ES absolute magnitude, and (ii) BAO measurements from \emph{DESI}, calibrated through the sound horizon derived from \emph{Planck} CMB data.

A natural route to alleviating this mismatch is to modify one of the two anchors—either reducing the sound horizon $r_d$~\cite{Hill:2020osr,Velten:2021cqj,Lee:2025yah,Mirpoorian:2025rfp,Jedamzik:2025cax,Mirpoorian:2024fka} or adjusting the SNe calibration. In this context, a violation of the DDR can be interpreted as an effective recalibration of either the SNe or BAO distance scale. The DDR holds under general conditions: in metric spacetime, photons propagate along null geodesics and their number is conserved. However, a variety of mechanisms can lead to DDR violations, including photon scattering by dust or gas~\cite{Menard:2009yb}, photon–axion conversion or decay~\cite{Avgoustidis:2010ju,Csaki:2001yk,Bassett:2003zw}, modified gravity~\cite{Cai:2015emx,Deffayet:2000pr}, and time variation of fundamental constants~\cite{Barrow:1999is,Lee:2021xwh,Goncalves:2019xtc}. These effects can be constrained by existing astrophysical and cosmological datasets~\cite{Zhou:2020moc,Qin:2021jqy,Bora:2021cjl,Mukherjee:2021kcu,Liu:2021fka,Renzi:2021xii,Tonghua:2023hdz,Qi:2024acx,Yang:2024icv,Tang:2024zkc,Jesus:2024nrl,Alfano:2025gie,Yang:2025qdg,Keil:2025ysb}.

Nevertheless, the DDR cannot directly modify BAO constraints, since BAO observables also depend on the longitudinal distance $D_H(z)$, which is unaffected by DDR transformations. Consequently, DDR violation is typically modeled as an effective modification of $D_L(z)$ through a constant offset in $\eta(z)$, reducing the SH0ES-calibrated $H_0$ toward the \emph{Planck} value. However, the Hubble tension is inherently multidimensional, involving not only $H_0$ but also the inferred matter density parameter $\Omega_m$, which must therefore be considered in any comprehensive resolution~\cite{Pedrotti:2024kpn}. A constant DDR offset does not alter the expansion history $E(z)$ and thus preserves the Pantheon-inferred $\Omega_m = 0.334 \pm 0.018$, which remains in tension with both CMB and weak-lensing results. Consequently, such a constant shift offers limited capacity to increase $H_0$ without worsening the $\Omega_m$ discrepancy.

To modify the SNe-calibrated $\Omega_{m,0}$, two main strategies exist: (i) altering the expansion history $E(z)$—for instance, through late-time dark energy dynamics—or (ii) introducing a redshift-dependent DDR violation, which modifies the $D_L$–$D_A$ mapping and forces $\Omega_m$ to readjust accordingly. The former has been widely explored~\cite{Teixeira:2025czm,Tutusaus:2023cms,Gomez-Valent:2023uof,Yang:2018qmz,Heisenberg:2022gqk,Zhou:2021xov,Escamilla:2023oce,Alestas:2021xes,Dai:2020rfo,Scherer:2025esj}; reducing $E(z \lesssim 0.5)$ to raise the BAO/CMB-inferred $H_0$ inevitably increases $\Omega_m$ (see detailed discussion in Section~\ref{sec:profile} ), thereby worsening the $S_8$–$\Omega_m$ tension. The latter—constructing a redshift-dependent $\eta(z)$—can align the SNe-inferred $\Omega_{m,0}$ with the \emph{Planck} value, but remains insufficient to fully resolve the $S_8$ discrepancy between CMB and weak-lensing data such as DES-Y3. Because the CMB tightly constrains the physical cold dark matter density $\omega_c$, lowering $\Omega_m = (\omega_b + \omega_c + \omega_\nu)/h^2$ necessarily requires an increase in $H_0$. Such simultaneous adjustments cannot arise from DDR violation alone, which tends to drive the SH0ES-calibrated $H_0$ back toward the \emph{Planck} baseline.

Based on these considerations, this work addresses two central questions:
\begin{enumerate}
	\item Can a redshift-dependent DDR violation mitigate the residual $\Omega_m$ tension—or equivalently, can the observed $\Omega_m$ discrepancy serve as indirect evidence for an evolving DDR? 
	\item To what extent are DDR violation and evolving dark energy degenerate, and can their interplay reconcile the multidimensional Hubble tension across CMB, BAO, and SNe datasets?
\end{enumerate}

To explore these questions, we employ the Fisher-bias formalism~\cite{Lee:2022gzh,Lee:2025yah}, which identifies minimal, data-driven deviations from the DDR while suppressing unphysical oscillatory modes. Unlike traditional Markov Chain Monte Carlo (MCMC) approaches, this method uses analytic gradients to infer the continuous function
$\eta(z) \equiv D_L(z)/[(1+z)^2 D_A(z)]$ implied by the data, offering both computational efficiency and physical transparency. We reconstruct $\eta(z)$ non-parametrically using Gaussian basis functions over $0 < z < 2$, aiming to identify cosmological shifts that reconcile CMB, BAO, and SNe calibrations.

In the following sections, we outline the Fisher-bias methodology, present the reconstructed $\eta(z)$ profiles from joint analyses of SNe, SH0ES, BAO, and CMB data, and discuss their implications for both the Hubble and $S_8$ tensions, as well as their possible connection to evolving dark energy.

This paper is organized as follows. 
Section~\ref{sec:method} introduces the Fisher-bias formalism and outlines its application to quantify parameter shifts arising from potential DDR violations.
Section~\ref{sec:reconstruction} presents the reconstructed $\eta(z)$ profiles and discusses their implications for reconciling early- and late-Universe measurements.
Section~\ref{sec:data} describes the observational datasets and covariance modeling adopted in our analysis.
Section~\ref{sec:mcmc} validates the Fisher-based reconstruction through full Markov Chain Monte Carlo (MCMC) analyses.
Finally, Section~\ref{sec:conclusion} summarizes our findings and highlights future directions.
\\


\begin{figure*}
	\begin{center}
		\includegraphics[scale = 0.48]{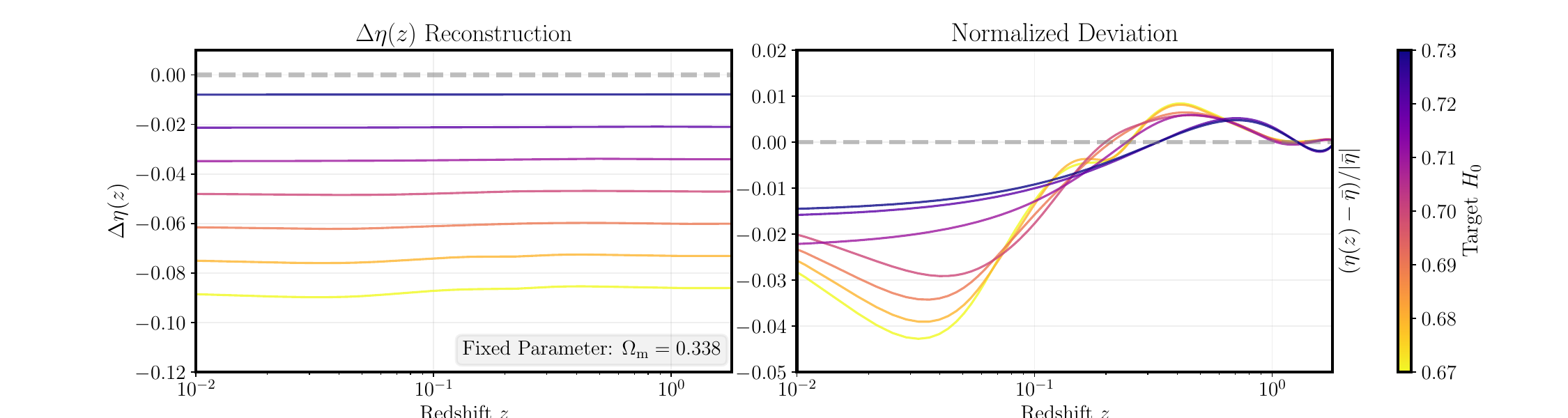}
		\includegraphics[scale = 0.48]{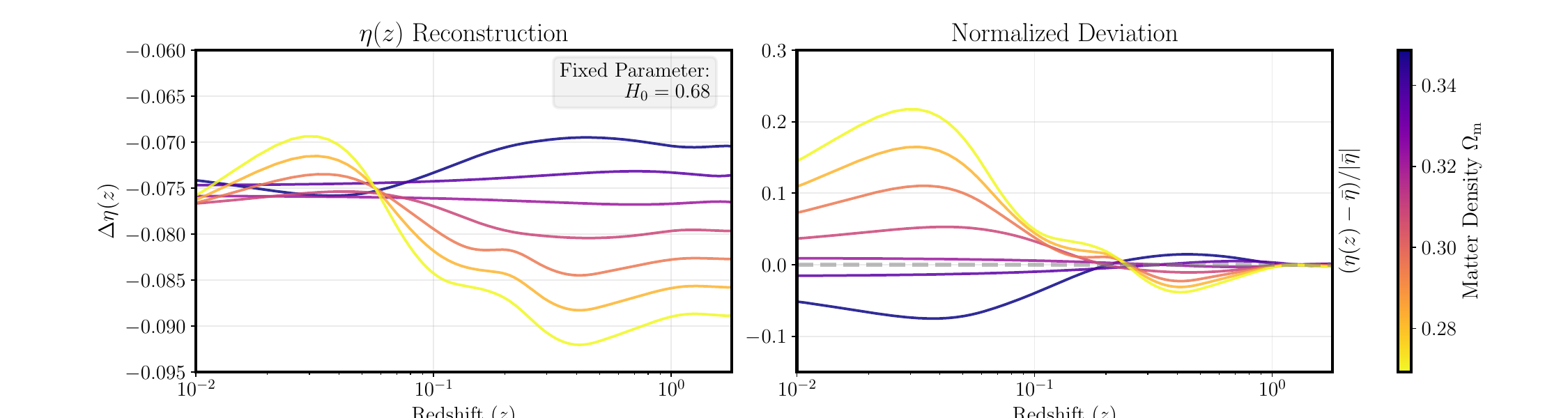}
	\end{center}
	\caption{Solutions for $\eta(z)$ profiles given target values of $H_0$ and $\Omega_m$. The vertical panels show solutions for: (a) fixed $\Omega_m = 0.338$ with different target $H_0$ values; (b) fixed $H_0 = 69.0$ with different target $\Omega_m$ values. The right panels display the reconstructed $\eta(z)$ profiles, while the left panels show the normalized deviation defined as $(\eta(z) - \bar{\eta})/|\bar{\eta}|$.}\label{fig:eta_h0_omm}
\end{figure*}

\section{Methodology}
\label{sec:method}

The distance--duality relation (DDR) connects the luminosity distance $D_L(z)$ and the angular--diameter distance $D_A(z)$ under very general physical conditions. In any metric theory of gravity where photons propagate along null geodesics and their number is conserved, the relation reads
\begin{equation}
	D_L(z) = (1+z)^2 D_A(z),
	\label{eq:ddr_def}
\end{equation}
which must hold in a statistically homogeneous and isotropic Universe. Neglecting any directional dependence, we focus on redshift--dependent deviations and define
\begin{equation}
	\eta(z)\equiv\frac{D_L(z)}{(1+z)^2 D_A(z)},
	\label{eq:eta_def}
\end{equation}
such that the standard DDR corresponds to $\eta(z)=1$.

To reconcile the SH0ES--anchored luminosity scale with the CMB--calibrated BAO distances without altering BAO observables, we treat DDR violation as an effective perturbation to $D_L(z)$ only. The deviation is expressed as
\begin{equation}
	\Delta\eta(z)\equiv\eta(z)-1=\sum_{i=1}^N c_i\,\phi_i(z),
\end{equation}
where $\{\phi_i(z)\}$ are localized basis functions. In this work we adopt two forms: (i) Gaussian kernels,
\begin{equation}
	\phi_i(z)=\exp\!\left[-\frac{(z-z_i)^2}{2\sigma^2}\right],
\end{equation}
which ensure smooth evolution, and (ii) top--hat windows,
\begin{equation}
	\phi_i(z)=
	\begin{cases}
		1, & |z-z_i|<\Delta z/2,\\
		0, & \text{otherwise},
	\end{cases}
\end{equation}
which allow for piecewise--constant deviations. These two forms bracket the range between continuous and sharply localized DDR deformations.

Our goal is to determine the minimal, redshift-dependent deformation $\eta(z)$ that shifts the supernova best-fit cosmological parameters $\vec{\Omega}_{\mathrm{BF}}$ to specified target values $\vec{\Omega}_{\mathrm{target}}$, while maintaining or improving the goodness of fit relative to the baseline $\Lambda$CDM model. The problem can be formulated as
\begin{equation}
	\begin{aligned}
		&\textrm{minimize}\quad 
		\int dz\, [\eta(z)-\bar{\eta}]^2 \\[3pt]
		&\textrm{subject to}\quad
		\begin{cases}
			\vec{\Omega}_{\rm BF}[\Delta\eta(z)] = \vec{\Omega}_{\rm target},\\[3pt]
			\Delta\chi^2_{\rm BF}[\Delta\eta(z)] \le 0,
		\end{cases}
	\end{aligned}
	\label{eq:optimization}
\end{equation}
where $\bar{\eta}$ is the mean of $\eta(z)$ over the reconstruction range. The quadratic penalty term suppresses spurious oscillatory behavior, ensuring that any redshift variation reflects physically required corrections rather than statistical noise.

We employ the Fisher--bias response formalism (see Refs.~\cite{Zhou:2025kws,Lee:2022gzh,Lee:2025yah}), which computes analytical response kernels describing the induced shifts in $(\vec{\Omega}_{\mathrm{BF}},\chi^2_{\mathrm{BF}})$ for a given perturbation $\Delta\eta(z)$. This method allows for a direct constrained minimization without full Monte Carlo sampling, yielding smooth, data-driven reconstructions while naturally suppressing noise-sensitive modes.

.

\section{Reconstruction of Late-Time Distance Duality}\label{sec:reconstruction}

All reconstructions are performed using the combined Pantheon+SH0ES likelihood with a Big Bang Nucleosynthesis (BBN) prior on $w_b$, and the baseline parameters are fixed to the best-fit $\Lambda$CDM solution of this combined likelihood. We then enforce external target values of either $H_0$ or $\Omega_m$ and reconstruct the minimal $\eta(z)$ deformation required to shift the Pantheon+SH0ES best-fit solution toward those targets without degrading the fit.

\subsection{Reconstructed $\eta(z)$ Profiles}\label{sec:profile}

The upper panel of Fig.~\ref{fig:eta_h0_omm} shows the case in which $\Omega_m$ is fixed at $0.338$, with the baseline corresponding to the Pantheon+SH0ES best fit. The reconstructed $\eta(z)$ profiles exhibit a monotonic dependence on the target $H_0$. For target values $H_0\gtrsim70~{\rm km\,s^{-1}\,Mpc^{-1}}$, $\eta(z)$ remains nearly constant, with fractional deviations satisfying 
\[
(\eta(z)-\hat{\eta})/\hat{\eta}\lesssim0.02\,.
\]
When the target $H_0$ is set to the \textit{Planck}+BAO best-fit value $H_0\simeq68~{\rm km\,s^{-1}\,Mpc^{-1}}$, we obtain an average normalization $\bar{\eta}\simeq0.925$, consistent with the result reported in Ref.~\cite{Teixeira:2025czm}.

The lower panel instead fixes $H_0=68~{\rm km\,s^{-1}\,Mpc^{-1}}$ while varying the target $\Omega_m$. In this configuration, both the amplitude and the shape of $\eta(z)$ depend sensitively on $\Omega_m$. For $z\gtrsim0.1$, $\eta(z)$ increases when $\Omega_m$ is raised above the Pantheon+SH0ES best-fit value, and decreases when $\Omega_m$ is lowered. When $\Omega_m$ is reduced to $\sim0.28$, the fractional deviation
\[
(\eta(z)-\hat{\eta})/\hat{\eta}
\]
reaches $\sim20\%$, indicating a highly significant redshift evolution. Assuming $\sigma_8$ is calibrated from the CMB and adopting weak-lensing constraints $S_8\approx0.77$ (implying $\Omega_m\simeq0.28$), the resulting discrepancy with Pantheon+SH0ES can thus be interpreted as indirect evidence for an evolving $\eta(z)$.

From a physical standpoint, only a limited number of mechanisms can alter the matter–density parameter inferred from Pantheon-like SNe samples within the framework of metric gravity, under the assumption of statistical isotropy. Any such change in the inferred $\Omega_m$ must originate from a modification of the functional form of the luminosity distance $D_L(z)$. This can occur either through a change in the background expansion rate $E(z)$, or via a redshift-dependent violation of the distance–duality relation, that is, $\eta(z)\neq1$.

A key implication of this framework is that suppressing the expansion rate 
$E(z)$ at $z \sim 0$--$0.5$---as typically realized by a late-time phantom transition 
around $z \sim 0.3$---to raise the $r_s$-calibrated BAO inference of $H_0$ 
inevitably drives the Pantheon+SH0ES--inferred matter density $\Omega_m$ to higher values. 
This shift amplifies the existing discrepancy with low-redshift structure-growth probes, 
quantified by 
$S_8 \equiv \sigma_8 \sqrt{\Omega_m / 0.3}$, 
where, for example, DES--Y3 finds 
$S_8 = 0.759^{+0.024}_{-0.021}$~\cite{DES:2021bvc}. 

From the perspective of the luminosity distance $D_L(z)$, 
a negative perturbation in the expansion rate 
($\delta E(z) < 0$ at $z \lesssim 0.5$) produces a positive response:
\begin{equation}
	\frac{\delta D_L}{D_L} \propto 
	- \int_0^z \frac{\delta E(z')}{E^2(z')} \, dz' > 0,
\end{equation}
since the integral $\int_0^z dz'/E(z')$ is dominated by low-redshift contributions. 
Given that Pantheon fixes the shape of $D_L(z)$ while SH0ES sets its absolute normalization, 
any excess in $D_L(z)$ must be compensated by an increase in $\Omega_m$, 
which raises $E(z)$ in the same range and restores the luminosity scale. 
Consequently, late-time modifications to $E(z)$ that enhance $H_0$ through BAO 
inherently push the inferred $\Omega_m$ upward, 
thereby worsening the combined $H_0$--$S_8$ tension unless an additional degree of freedom, 
such as an evolving $\eta(z)$, is introduced. 

In contrast, a time-dependent $\eta(z)$ offers an independent mechanism 
to displace the SNe-inferred $\Omega_m$ 
(see the lower panel of Fig.~\ref{fig:eta_h0_omm}), 
thereby alleviating the luminosity-scale inconsistency 
without simultaneously aggravating the matter-density discrepancy. 

Taken together, these reconstructed profiles demonstrate that 
the required deformation of $\eta(z)$ is jointly and sensitively controlled 
by the target values of $H_0$ and $\Omega_m$, 
emphasizing that the late-- versus early--Universe calibration problem 
is intrinsically multidimensional rather than reducible 
to a single-parameter shift in $H_0$.

\begin{figure}
	\begin{center}
		\includegraphics[scale = 0.34]{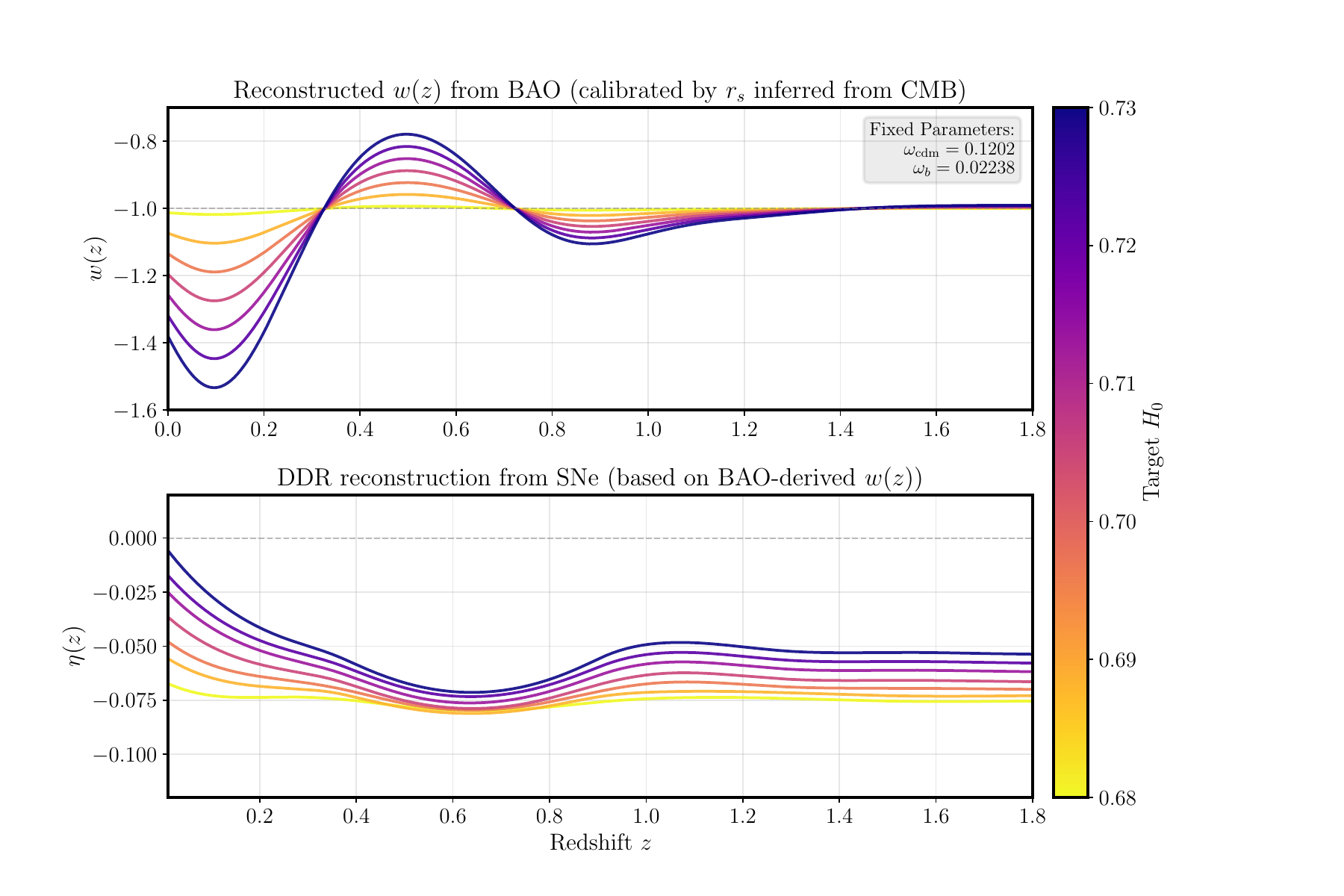}
	\end{center}
	\caption{Reconstructed $w(z)$ and $\eta(z)$ profiles based on BAO data. The upper panels show the $w(z)$ profile reconstructed using BAO and $r_s$ derived from the CMB likelihood, with $\omega_c = 0.1202$ and a fixed target $H_0$. The lower panels display the corresponding $\eta(z)$ profiles obtained using the Fisher-bias method, with the same $\omega_c$ and $H_0$ target. All solutions preserve the Planck $\Lambda$CDM best-fit parameters.		
	}\label{fig:wz_ddr_combined}
\end{figure}

\begin{figure}
	\begin{center}
		\includegraphics[scale = 0.42]{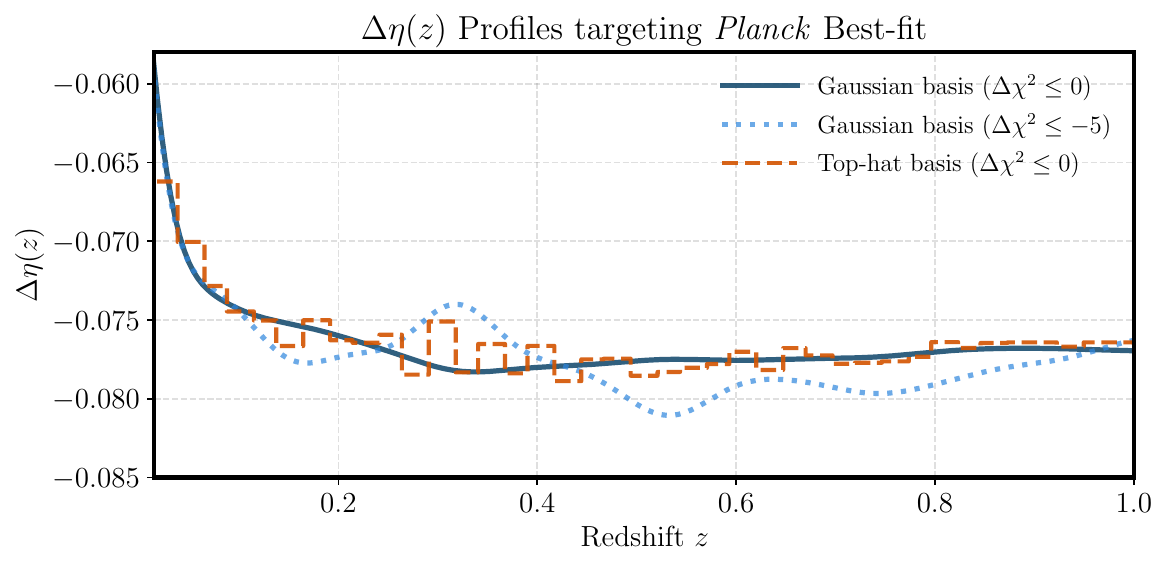}
	\end{center}
	\caption{Comparison of $\eta(z)$ profiles reconstructed using Gaussian and Top-hat basis functions. The solid line represents the result obtained using Gaussian basis functions (M2-Timdep model), while the dashed line corresponds to the Top-hat basis result. Both profiles are derived from the Pantheon+SH0ES likelihood, with target parameters fixed to the best-fit values from the Planck + BAO $\Lambda$CDM model ($\omega_c = 0.1202$, $H_0 = 0.684$).}\label{fig:eta_gauss_tophat}
\end{figure}

\subsection{Joint $\eta(z)$--$H(z)$ Deformation and the $\omega_c$--$H_0$ Coupling}


Modifying $\eta(z)$ alone can indeed reduce the Pantheon-inferred $\Omega_m$, 
but this deformation does not propagate to the BAO+CMB sector. 
Given that $\omega_c \equiv \Omega_c h^2$ and $\omega_b \equiv \Omega_b h^2$ are tightly constrained 
by the CMB and BBN likelihoods, increasing $H_0$ inevitably lowers
\[
\Omega_m = \Omega_c + \Omega_b ,
\]
which accounts for the $S_8$ tension between CMB and weak-lensing measurements. 
Therefore, reducing the CMB-inferred $\Omega_m$ requires a modification of the expansion history $E(z)$ 
to raise the $H_0$ value jointly inferred by CMB+BAO. Although such a modification tends 
to increase the Pantheon-inferred $\Omega_m$, this uplift can be counterbalanced 
by a time-dependent $\eta(z)$ deformation.

Addressing the $S_8$ tension further necessitates a reduction in $\Omega_m$, 
which in turn requires a coupled deformation of both $\eta(z)$ and $H(z)$ rather than $\eta(z)$ alone.

Our strategy proceeds as follows. Using BAO and $r_s$ derived from the CMB likelihood, 
we first fix $\omega_c = 0.1202$ and impose a target $H_0$. 
We then obtain an optimal $w(z)$ profile via the Fisher-bias optimization method. 
This model serves as the new baseline, with the Pantheon+SH0ES best-fit values 
within this model as reference parameters. Subsequently, we apply the Fisher-bias method again, 
fixing $\omega_c = 0.1202$ and the same $H_0$ target, 
to reconstruct the corresponding $\eta(z)$ profile.

Figure~\ref{fig:wz_ddr_combined} illustrates different $H_0$ targets in color: 
the upper panel shows the reconstructed $w(z)$, 
and the lower panel shows the corresponding $\eta(z)$. 
As the equation of state exhibits a phantom crossing around $z\sim0.3$, 
evolving from $w<-1$ at $z<0.3$ to $w>-1$ thereafter, 
$\eta(z)$ develops a pronounced dip. 
Higher $H_0$ targets produce deeper dips, with the minimum located near $z\sim0.5$. 
The profile mildly rebounds near $z\sim1$ and becomes approximately constant for $z>1$. 
This behavior is consistent with the independently reconstructed $\eta(z)$-only deformation, 
confirming that reducing the Pantheon-inferred $\Omega_m$ requires a declining $\eta(z)$ profile.

Physically, the dip arises from the response of the luminosity distance to a reduced matter density. 
In a flat cosmology with fixed $H_0$, decreasing $\Omega_m$ lowers the background expansion rate in the matter-dominated regime:
\[
E(z) = \frac{H(z)}{H_0} = \sqrt{\Omega_m(1+z)^3 + \Omega_\Lambda}, \quad \partial_{\Omega_m} E(z) > 0.
\]
This reduction increases the luminosity distance,
\[
D_L(z) \propto \int_0^z \frac{dz'}{E(z')}, \quad \partial_{\Omega_m} D_L(z) < 0 \quad \text{for } z \sim 0.3\text{--}1.
\]
Since Pantheon fixes the distances, this increase must be compensated by a suppressive $\eta(z)$ deformation 
over the same redshift range. The observed dip is therefore the minimal compensating mode needed 
to offset the $D_L$ response to a lower $\Omega_m$, 
while preserving the low-$z$ normalization (SH0ES calibration) and high-$z$ asymptotic behavior.


\section{Statistical Methodology and Datasets}\label{sec:data}

We implement the distance-duality relation (DDR) scenarios by modifying the publicly available Einstein-Boltzmann code \textsc{CLASS} \cite{Lesgourgues:2011re,Blas:2011rf}. Our analysis considers three distinct models:

\begin{itemize}
	\item[(i)] \textbf{M1-Const: Constant DDR violation.}
	This model introduces a constant deviation from the standard DDR, fixed at $\eta(z) = 0.925$ (corresponding to $\Delta\eta = -0.075$).
	
	\item[(ii)] \textbf{M2-TimeDep: Time-dependent DDR.}  
	This model incorporates a redshift-dependent $\eta(z)$ profile reconstructed via our Fisher-bias optimization. The target parameters are set to the \textit{Planck} $\Lambda$CDM + BAO best-fit values ($\omega_c = 0.1202$, $h = 0.684$), while the baseline adopts the $\Lambda$CDM best-fit from the combined Pantheon+ and SH0ES likelihood. The resulting $\eta(z)$ profiles—shown in Fig.~\ref{fig:eta_gauss_tophat}—are reconstructed using  Gaussian basis functions\footnote{In subsequent MCMC analyses, we adopt the Gaussian-reconstructed profile with a full width at half maximum (FWHM) of 3, optimized for $\Delta \chi^2 \leq -5$ relative to the baseline $\Lambda$CDM fit.}.
	
	\item[(iii)] \textbf{M3-Hybrid: Hybrid model.}  
	This model jointly incorporates an optimized $w(z)$ profile—targeting $H_0 = 71$ and $\omega_c = 0.1202$ using BAO calibrated with the CMB-inferred sound horizon $r_s$—together with a corresponding $\eta(z)$ profile reconstructed from the Pantheon+ and SH0ES likelihood under the same parameter targets.
	
\end{itemize}

We perform Markov Chain Monte Carlo (MCMC) analyses using the public code \textsc{MontePython}\footnote{\href{https://github.com/brinckmann/montepython_public}{https://github.com/brinckmann/montepython\_public}}~\cite{Brinckmann:2018cvx}, adopting the Gelman-Rubin convergence criterion \cite{10.1214/ss/1177011136} $R-1 < 0.05$. All statistical plots are generated with the \textsc{GetDist} package \cite{Lewis:2019xzd}. We utilize multiple cosmological probes spanning different redshifts and physical scales, combining DESI BAO, Pantheon+ SNe Ia, cosmic chronometers, and \textit{Planck} distance priors, together with their full covariance matrices.

\subsection{DESI BAO (DR2)}
\label{subsec:desi_dr2}

The Dark Energy Spectroscopic Instrument (DESI) Data Release 2 \cite{DESI:2025zgx} provides BAO measurements using 14 million extragalactic objects across four distinct tracer classes. The BAO measurements are reported in nine redshift bins spanning $0.295 \leq z \leq 2.330$ as:
\begin{equation}
	\left( \frac{D_M(z)}{r_d}, \frac{D_H(z)}{r_d} \right),
\end{equation}
where $D_M(z)$ is the transverse comoving distance, $D_H(z) = c/H(z)$ is the Hubble distance, and $r_d$ is the sound horizon at the drag epoch. We incorporate the full non-diagonal covariance matrix accounting for cross-correlations between redshift bins and tracer types.

\subsection{Pantheon+SH0ES Supernovae}
\label{subsec:panthon_plus}

The Pantheon+ compilation~\cite{Scolnic:2021amr,Brout:2022vxf} provides distance moduli derived from 1701 light curves of 1550 spectroscopically confirmed Type~Ia supernovae (SNe~Ia) spanning the redshift range $0.001 < z < 2.26$. To incorporate local distance-scale information, we include the SH0ES calibration through the full PantheonPlus+SH0ES likelihood( see Ref.~\cite{Poulin:2024ken}).

\subsection{Cosmic Chronometers}
\label{subsec:cc}
Cosmic chronometer measurements directly constrain the Hubble parameter through differential aging of passively evolving galaxies. We utilize 32 measurements\cite{Zhang:2012mp,Simon:2004tf,Moresco:2012jh,Moresco:2015cya,Ratsimbazafy:2017vga,Stern:2009ep,Borghi:2021rft} summarized in Ref. \cite{Wu:2025wyk} spanning $0.07 < z < 1.97$. The covariance matrix accounts for systematic uncertainties in stellar population synthesis models.

\subsection{Planck datasets}
\label{subsec:planck_full}
We consider the CMB distance prior derived from final Planck 2018 release\cite{Chen:2018dbv} in Fisher-Bias analysis. These priors include the shift parameter $\mathcal{R}$, the acoustic scale $\ell_A$, and the baryon density $\Omega_b h^2$.
For MCMC validation of solutions,  we employ the \emph{Planck} 2018 low-$\ell$ TT+EE and \emph{Planck} 2018 high-$\ell$ TT+TE+EE temperature and polarization power spectrum \cite{Planck:2019nip, Planck:2018lbu}. 

\begin{figure*}
	\begin{center}
		\includegraphics[scale = 0.28]{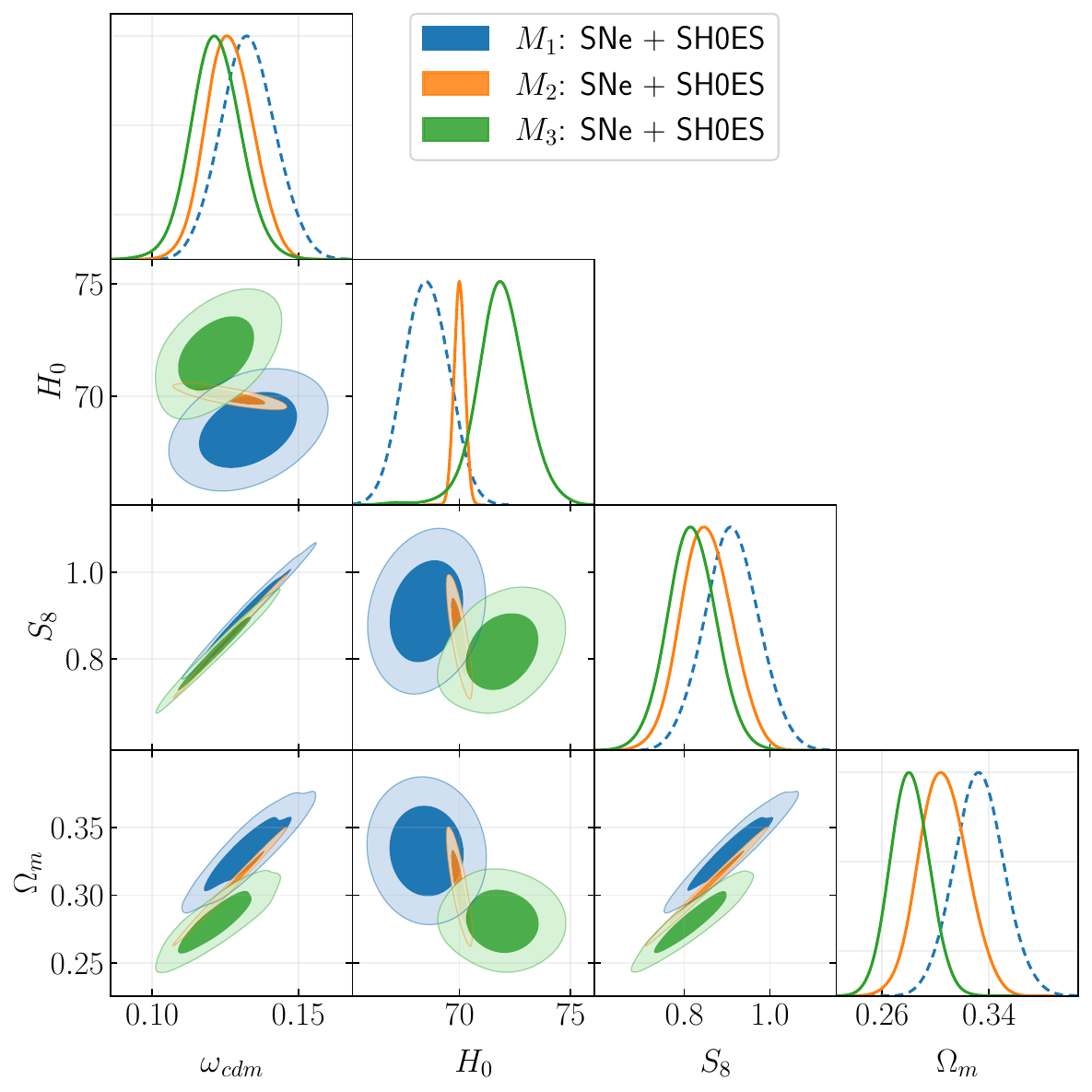}
		\includegraphics[scale = 0.28]{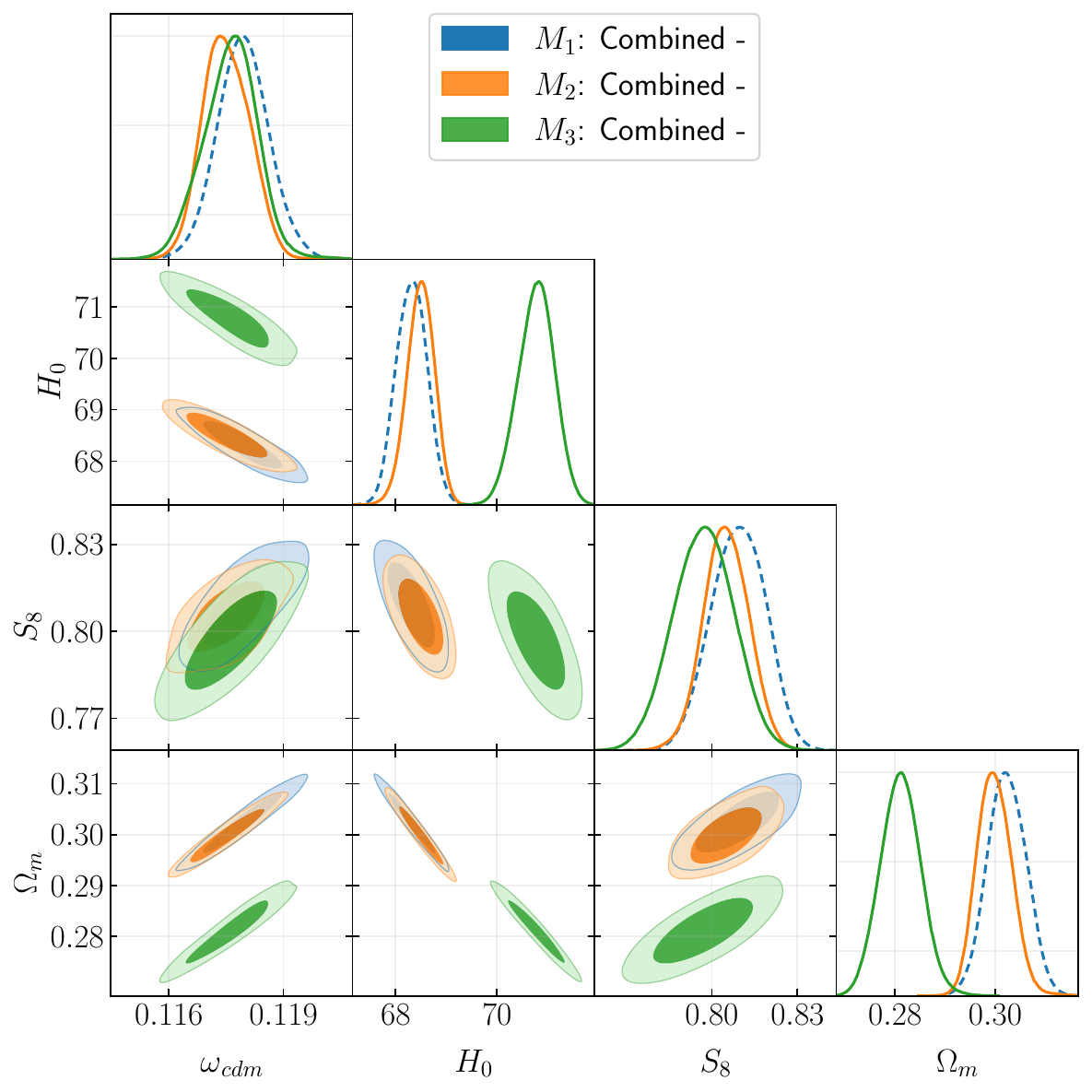}
		\includegraphics[scale = 0.28]{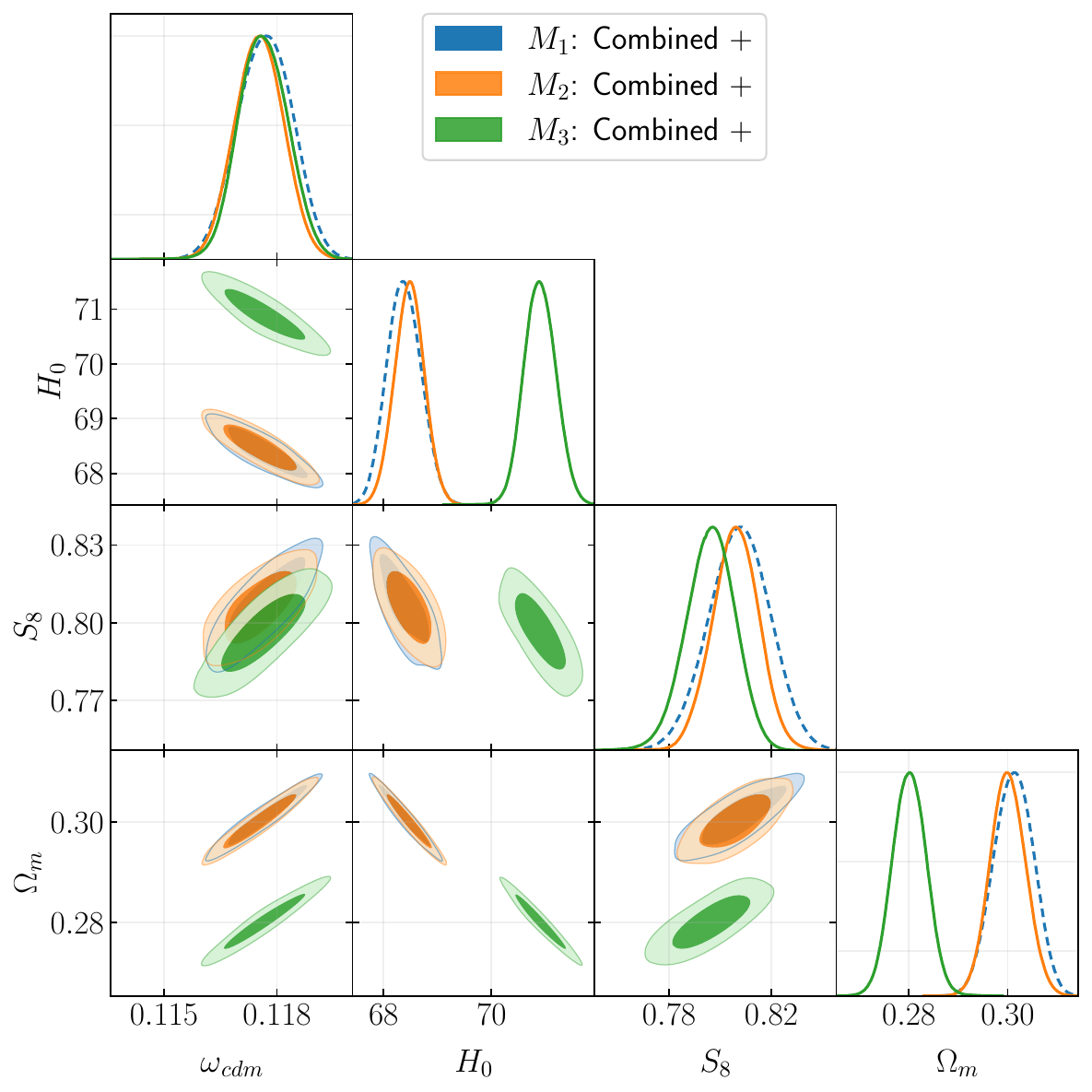}
	\end{center}
	\caption{Constraints on cosmological parameters for the three compared models (M1-Const, M2-TimeDep, M2-Hybrid) from different data combinations. The ``M$-$'' label denotes data combinations excluding the SH0ES prior, specifically including CC + BAO + SNe + \emph{Planck} likelihoods, while ``M$+$'' indicates the same combinations with the SH0ES calibration incorporated.
	}\label{fig:compare_contour}
\end{figure*}


\begin{table*}[t]
	\renewcommand{\arraystretch}{1.4}
	\setlength{\tabcolsep}{5pt}
	\centering
	\caption{
		Cosmological parameter constraints for the M1--Const, M2--TimDep, M3--Hybrid, and $\Lambda$CDM models across different data combinations. 
		All values represent the mean $\pm1\sigma$ confidence intervals. 
		Here, the ``M$-$'' label denotes data combinations excluding the SH0ES prior, specifically including CC + BAO + SNe + \emph{Planck} likelihoods, 
		while ``M$+$'' indicates the same combinations with the SH0ES calibration incorporated.
	}
	\label{tab:combined_results}
	\begin{tabular}{|l|l|cccccc|c|}
		\hline
		\hline
		Dataset & Model & $M_{\mathrm{B}}$ & $\omega_{\mathrm{cdm}}$ & $H_0$ & $S_8$ & $\Omega_{m,0}$ & $\chi^2$ \\
		\hline
		\hline
		\multirow{4}{*}{SNe+SH0ES} 
		& M1 & $-19.233\pm 0.029$ & $0.1328\pm 0.0093 $ & $68.5\pm 0.9$ & $0.911\pm 0.064$ & $0.333\pm 0.019$ & 1286.37 \\
		& M2 & $-19.2082\pm 0.0072$ & $0.1265\pm 0.0080 $ & $69.4\pm 0.24 $ & $0.852\pm 0.058   $ & $0.306\pm 0.018 $ & 1279.58 \\
		& M3 & $-19.236\pm 0.034$ & $0.1218\pm 0.0086$ & $71.9\pm 1.1 $ & $0.817\pm 0.057 $ & $0.281\pm 0.015 $ & 1282.94 \\
		& $\Lambda$CDM & $-19.228\pm 0.010$ & $0.161\pm 0.010$ & $74.19^{+0.42}_{-0.50}$ & $1.047\pm 0.069$ & $0.335\pm 0.020$ & 1286.38 \\
		\hline
		\multirow{4}{*}{Combined -} 
		& M1 & $-19.2442\pm 0.0092$ & $0.1179\pm 0.0007$ & $68.32\pm 0.30$ & $0.8093\pm 0.0094$ & $0.3023\pm 0.0038$ & 2447.68 \\
		& M2 & $-19.247\pm 0.010 $ & $0.1176\pm 0.0007$ & $68.43\pm 0.32  $ & $0.8072\pm 0.0093$ & $0.3006\pm 0.0040$ & 2436.46 \\
		& M3 & $-19.2500^{+0.0077}_{-0.010}$ & $0.1178^{+0.0007}_{-0.0006}$ & $71.37^{+0.26}_{-0.36} $ & $0.796\pm 0.011 $ & $0.2767^{+0.0038}_{-0.0031}$ & 2436.98 \\
		& $\Lambda$CDM & $-19.4009^{+0.0060}_{-0.0069}$ & $0.1172\pm 0.0006$ & $68.76^{+0.17}_{-0.20}$ & $0.8028^{+0.0076}_{-0.0089}$ & $0.2973\pm 0.0029 $ & 2448.8 \\
		\hline
		\multirow{4}{*}{Combined +} 
		& M1 & $-19.2463\pm 0.0090$ & $0.1177\pm 0.0007 $ & $68.38\pm 0.28$ & $0.808\pm 0.010 $ & $0.3012\pm 0.0036$ & 2324.62 \\
		& M2 & $-19.2471\pm 0.0072 $ & $0.1175\pm 0.0006$ & $68.53\pm 0.24$  & $0.8044\pm 0.0083 $ & $0.2995\pm 0.0030$ & 2313.14 \\
		& M3 & $-19.2492\pm 0.0088$ & $0.1176^{+0.0007}_{-0.0006}$ & $71.48^{+0.30}_{-0.35}$ & $0.796\pm 0.012$ & $0.2755^{+0.0038}_{-0.0034}$ & 2315.74 \\
		& $\Lambda$CDM & $-19.3982\pm 0.0081 $ & $0.1169\pm 0.0006$ & $68.82\pm 0.26$ & $0.80^{+0.0092}_{-0.0081}$ & $0.2961\pm 0.0032$ & 2356.26 \\
		\hline
	\end{tabular}
\end{table*}


\section{MCMC Analysis}
\label{sec:mcmc}

To validate the $\eta(z)$ profiles reconstructed via the Fisher-bias approach, we perform full Markov Chain Monte Carlo (MCMC) analyses for three representative models introduced in Section~\ref{sec:data}, namely M1-Const, M2-TimDep, and M3-Hybrid. Each reconstructed $\eta(z)$ profile is implemented in \href{http://class-code.net/}{CLASS} and held fixed at its Fisher-bias–derived form during sampling.

For clarity, in the following results the M$-$'' label denotes data combinations excluding the SH0ES prior, specifically including CC + BAO + SNe + \emph{Planck} likelihoods, while M$+$'' indicates the same combinations with the SH0ES calibration incorporated.

\begin{figure*}
	\begin{center}
		\includegraphics[scale = 0.28]{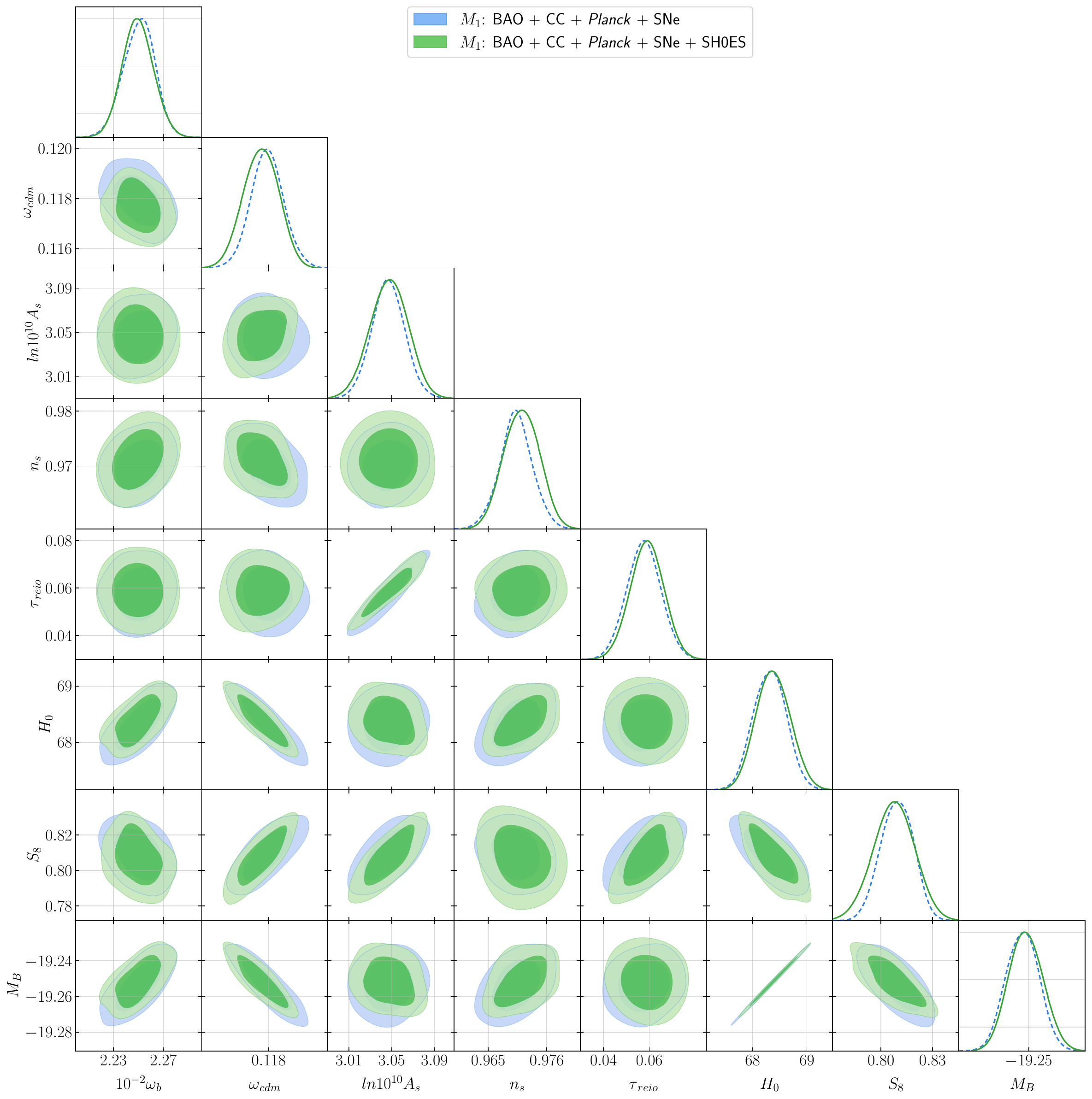}
		\includegraphics[scale = 0.28]{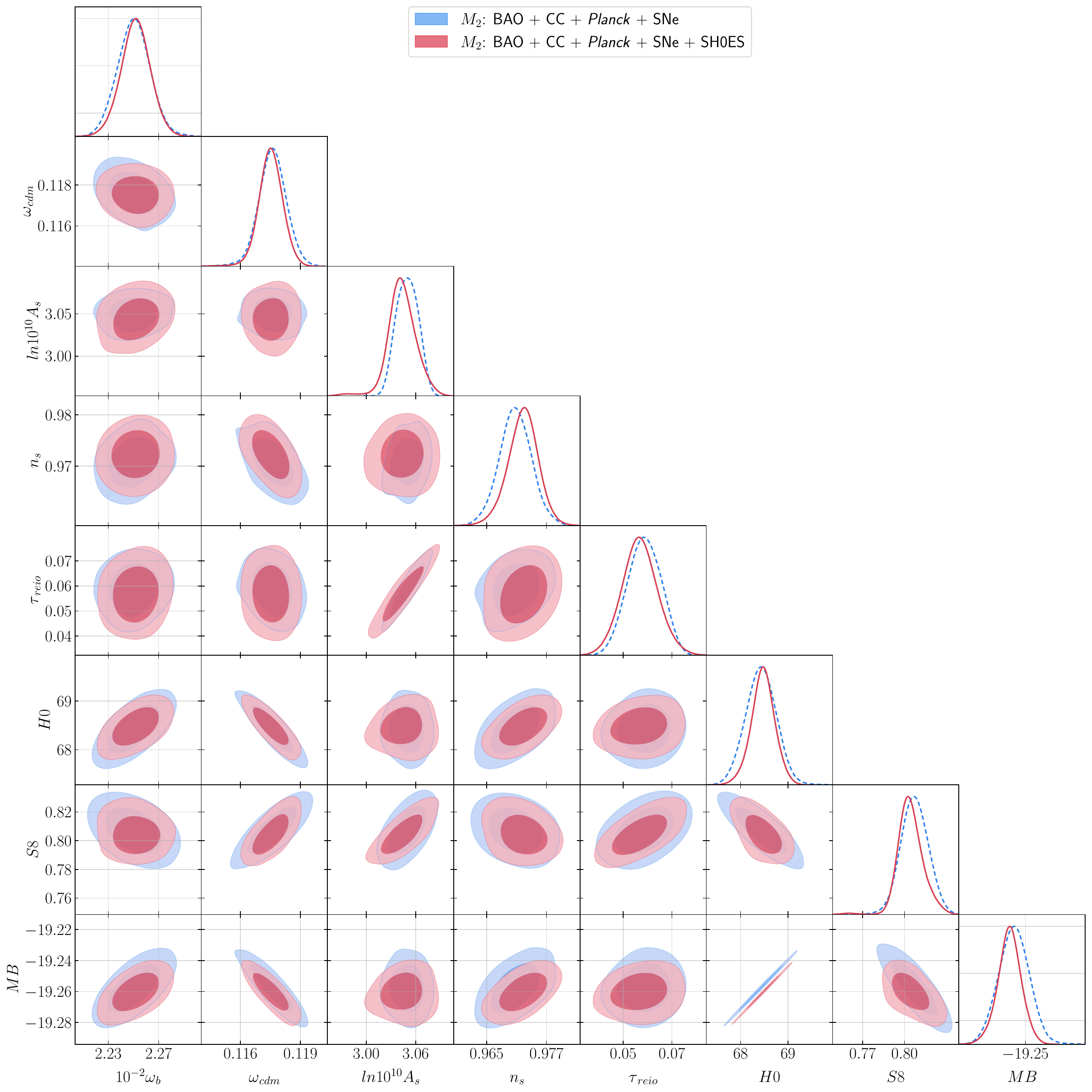}
	\end{center}
	\caption{Cosmological parameter constraints from different data combinations. 
		The upper panel shows the M1-Const model targeting ($\eta = 0.925$), while the lower panel displays the M2-TimeDep model. For each model, we show two data combinations:  
		(i) BAO + CC + \emph{Planck} + SNe , and 
		(ii) BAO + CC + \emph{Planck} + SNe + SH0ES. 
		All contours represent $1\sigma$ and $2\sigma$ confidence regions.
	}\label{fig:M1_M2_contour2}
\end{figure*}

\begin{figure*}
	\begin{center}
		\includegraphics[scale = 0.28]{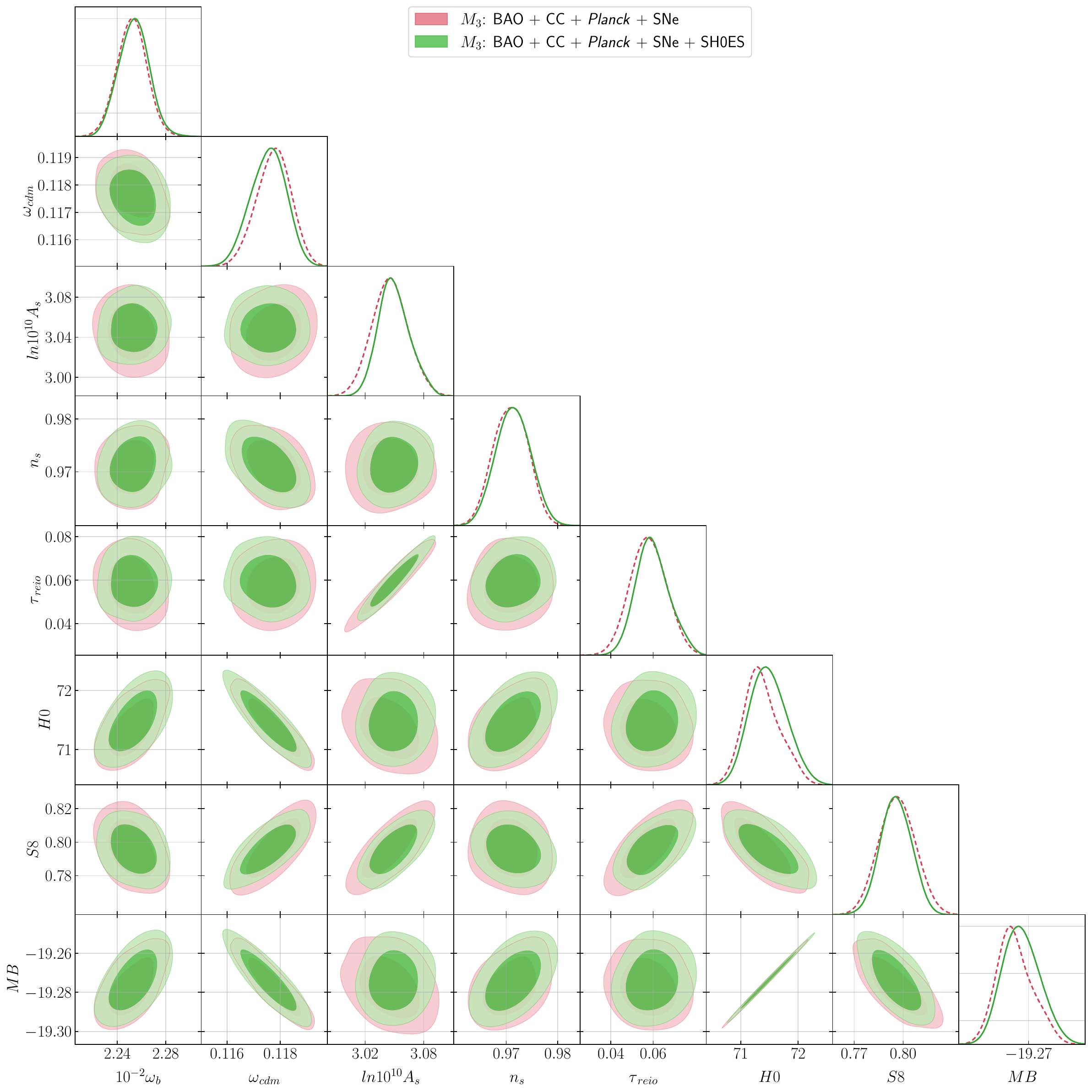}
	\end{center}
	\caption{ Cosmological parameter constraints for the M3-Hybrid model from different data combinations. we show two data combinations:  
		(i) BAO + CC + \emph{Planck} + SNe , and 
		(ii) BAO + CC + \emph{Planck} + SNe + SH0ES. 
		All contours represent $1\sigma$ and $2\sigma$ confidence regions.
	}\label{fig:M3_contour}
\end{figure*}

As summarized in Table~\ref{tab:combined_results} and illustrated in Figs.~\ref{fig:compare_contour}, \ref{fig:M1_M2_contour2} and \ref{fig:M3_contour}, the MCMC results consistently recover the target cosmological parameters within the $2\sigma$ confidence intervals. For the M2-TimDep model—optimized for the Pantheon SNe + SH0ES dataset with target values $H_0 = 68.4$ km/s/Mpc and $\Omega_c = 0.1202$—the MCMC inference yields $H_0 = 69.4 \pm 0.24$ km/s/Mpc and $\omega_{\rm cdm} = 0.1218 \pm 0.0086$ for M$+$, and $H_0 = 68.53 \pm 0.24$ km/s/Mpc and $\omega_{\rm cdm} = 0.1175 \pm 0.0006$ for M$-$, demonstrating close agreement with the intended parameter targets.

\subsection{Model Contrast}

In comparison with the $\Lambda$CDM model, the M1-Const model yields nearly identical matter densities when refitting the Pantheon + SH0ES dataset, giving $\Omega_m = 0.333 \pm 0.019$ and $0.335 \pm 0.020$, respectively. This confirms that a constant distance-duality relation (DDR) does not alter the Pantheon inference of $\Omega_m$. From the perspective of goodness-of-fit, when the SH0ES likelihood is excluded (combined$-$ case), the two models exhibit nearly identical $\chi^2$ values, with $\Delta\chi^2 \simeq -1$. Once the SH0ES prior is included, the M1-Const model achieves a substantially better fit, with $\Delta\chi^2 \simeq -30$, consistent with the findings of Ref.~\cite{Teixeira:2025czm}. This indicates that a constant DDR can effectively alleviate the apparent $H_0$ tension.

Nevertheless, for the M1-Const model, the $S_8$--$\Omega_m$ relation inferred from different datasets shows a significant internal discrepancy: for the SNe + SH0ES case, $S_8 = 0.911 \pm 0.064$ and $\Omega_m = 0.333 \pm 0.019$, while for the combined$-$ case, $S_8 = 0.8093 \pm 0.0094$ and $\Omega_m = 0.3023 \pm 0.0038$. The internal tension between these two determinations corresponds to approximately $1.6\sigma$ in $\Omega_m$ and $1.6\sigma$ in $S_8$.

When a time-varying DDR is introduced in the M2-TimDep model—characterized by a pronounced dip in the $\eta(z)$ profile around $z\simeq0.1$—the discrepancy between datasets is markedly reduced. For SNe + SH0ES, we obtain $S_8 = 0.852 \pm 0.058$ and $\Omega_m = 0.306 \pm 0.018$, while for the combined$-$ case, $S_8 = 0.8072 \pm 0.0094$ and $\Omega_m = 0.3006 \pm 0.0038$, corresponding to only $\sim 0.8\sigma$ in $\Omega_m$ and $\sim 0.7\sigma$ in $S_8$. For the hybrid M3-Hybrid model, the residual tension is further reduced.

In terms of overall likelihood, both M2-TimDep and M3-Hybrid achieve $\Delta\chi^2 \simeq -10$ relative to $\Lambda$CDM in the combined$-$ configuration. This improvement arises primarily from reduced cross-dataset tension and an improved fit to the Pantheon data itself (optimized for $\Delta\chi^2 < -5$). In the combined$+$ configuration (i.e., when the SH0ES prior is included), the improvement becomes much more pronounced, reaching $\Delta\chi^2 \simeq -40$. Hence, the residual $S_8$--$\Omega_m$ discrepancy could be interpreted as an indication of a time-dependent DDR.

Finally, the inferred absolute magnitude $M_B$ remains highly consistent across all data combinations in the M1-Const, M2-TimDep, and M3-Hybrid models, with differences well below $1\sigma$. This stability demonstrates the internal coherence of the reconstructions and confirms that the inferred parameters remain robust regardless of whether the SH0ES prior is included.

\subsection{Degeneracy between DDR Violation and Evolving Dark Energy}

Our analysis reveals that the M2-TimDep (pure time-dependent $\eta(z)$) and M3-Hybrid (hybrid $\eta(z)$ + $w(z)$) models yield nearly identical constraints on key parameters—such as $\omega_{\mathrm{cdm}}$ and $M_{\mathrm{B}}$—within $1\sigma$ uncertainties when fitted to the same combination of CMB, BAO, and SNe~Ia data. This indicates a significant degeneracy between the two types of extensions: a phantom transition in $w(z)$ at late times can be effectively counterbalanced by a dip in $\eta(z)$ at $z \sim 0.1$--$0.5$, leading to similar predictions for the observed luminosity distances.

This degeneracy arises because a phantom-like equation of state ($w<-1$) at low redshift suppresses the expansion rate $E(z)$, which in turn increases the luminosity distance $D_L(z)$. This enhancement can be offset by a decrease in $\eta(z)$, which multiplicatively reduces $D_L(z)$. As a result, the two competing effects can cancel out in the predicted distances, making it difficult to distinguish between them using $E(z)$-sensitive observables alone. In the absence of additional measurements that directly and independently probe either the expansion history $H(z)$ or the distance-duality relation itself, current datasets cannot unambiguously differentiate between these two physical mechanisms.

To break this degeneracy, independent measurements of the expansion history are essential. Cosmic chronometers (CC), which provide direct estimates of $H(z)$ from the differential ages of passively evolving galaxies, offer a cosmology-independent probe of the expansion rate. Similarly, the radial BAO observable $D_H(z) \equiv c/H(z)$—when combined with a robust sound-horizon prior $r_d$—provides another direct constraint on $H(z)$. Incorporating such data can help isolate the respective contributions of $H(z)$ and $\eta(z)$ to the observed distances, thereby clarifying whether the apparent anomalies arise from a genuine violation of distance duality, an evolving dark energy component, or a combination of both.

\subsection{Resolution of the Multi-dimensional Tensions}

Once $\Omega_{m,0}$ is calibrated by SNe~Ia observations, any attempt to raise $H_0$ generally requires an increase in $\omega_c$, thereby exacerbating the tension with low-redshift large-scale structure probes. As discussed earlier, employing a redshift-dependent $\eta(z)$ in the M2-TimDep model can align the SNe-inferred $\Omega_{m,0}$ with the CMB-preferred value, but it remains insufficient to fully resolve the $S_8$ tension between the CMB and weak-lensing surveys such as DES-Y3.

To further mitigate the residual $S_8$ and $H_0$ discrepancies, we introduce the hybrid M3-Hybrid model, which combines a phantom-like dark energy transition with a time-dependent DDR. In this scenario, the inferred Hubble constant is consistently higher than that in M1-Const, M2-TimDep, and $\Lambda$CDM, while the corresponding matter density $\Omega_{m,0}$ is systematically lower than in all three. Remarkably, across all dataset combinations—whether or not the SH0ES prior is included—the inferred $\Omega_{m,0}$ remains stable at approximately $0.28$, indicating the absence of any internal tension among different observational probes. The corresponding amplitude of matter fluctuations is $S_8 = 0.796 \pm 0.012$, which effectively alleviates the tension with the DES-Y3 result.

These findings suggest that the M3-Hybrid model provides a more coherent reconciliation of the multi-faceted cosmological tensions—particularly the simultaneous mitigation of the $H_0$ and $S_8$ discrepancies—outperforming both the M2-TimDep and M1-Const models.

\section{Conclusion}\label{sec:conclusion}

In this work, we have investigated whether violations of the distance-duality relation (DDR), parameterized by $\eta(z) = D_L(z)/[(1+z)^2 D_A(z)]$, can help alleviate the multidimensional tensions among cosmological parameters inferred from CMB, BAO, and Type~Ia supernova (SNe~Ia) observations. Using the Fisher-bias formalism—validated through full Markov Chain Monte Carlo (MCMC) analyses—we reconstructed minimal, data-driven $\eta(z)$ profiles that capture the late-time deviations required to reconcile early- and late-Universe calibrations.

Our analysis shows that a constant DDR offset (M1-Const model) can partially mitigate the apparent $H_0$ discrepancy, yielding $\Delta\chi^2 \simeq -30$ when the SH0ES prior is included. However, it leaves the Pantheon-inferred matter density parameter $\Omega_m = 0.334 \pm 0.018$ nearly unchanged, thereby maintaining its inconsistency with the \emph{Planck} best-fit $\Lambda$CDM model and weak-lensing surveys that prefer lower $\Omega_m$ and $S_8$ values. Allowing the DDR to evolve with time (M2-TimeDep model), featuring a dip in $\eta(z)$ around $z \sim 0.1$–$0.5$, significantly improves the global fit, with $\Delta\chi^2 \simeq -10$ relative to $\Lambda$CDM or M1-Const when SH0ES is excluded, while reducing the internal $\Omega_m$ tension from $1.6\sigma$ to below $1\sigma$. This scenario aligns the SNe-calibrated $\Omega_{m,0}$ with the CMB-preferred value, though it remains unable to fully resolve the $S_8$ discrepancy due to the stringent CMB constraints on $\omega_c$. 

It is worth noting that reconstructions of $H(z<2)$ preserving the standard DDR—aimed solely at reconciling the $H_0$ discrepancy between SH0ES and $r_s$-calibrated BAO—inevitably exacerbate the $\Omega_m$ inconsistency. In this context, the observed $\Omega_m$ discrepancy may be interpreted as indirect evidence for a time-varying DDR, although the underlying physical mechanism remains ambiguous within current observational precision.

The hybrid model (M3-Hybrid), which combines a phantom-like dark energy transition with a time-dependent DDR, provides the most consistent global reconciliation across all dataset combinations. In this framework, the inferred $\Omega_{m,0}$ remains stable around $\sim0.28$ across different dataset combinations, with $S_8 = 0.796 \pm 0.012$ when all datasets are combined, effectively reducing the tension with DES-Y3 measurements to below $2\sigma$. These results suggest that a mild DDR violation, coupled with evolving dark energy, offers a coherent and physically motivated pathway toward jointly addressing the $H_0$ and $S_8$ tensions.

We further identify a strong degeneracy between DDR violation and dynamical dark energy. A phantom transition in the dark energy equation of state $w(z)$ can mimic the effects of a redshift-dependent $\eta(z)$, leading to nearly identical predictions for luminosity distances and cosmological parameter constraints. This degeneracy highlights the challenge of disentangling geometric modifications to the distance relation from dynamical changes in the expansion history using current datasets. Breaking this degeneracy requires independent probes of the expansion rate $H(z)$. Cosmic chronometers and the radial BAO observable $D_H(z) = c/H(z)$, when combined with precise sound-horizon measurements, offer a promising route to isolate potential DDR-related effects.

Our analysis adopts a phenomenological perspective, and we defer the construction of a concrete physical mechanism capable of generating the reconstructed DDR deviations to future studies.


\begin{acknowledgments}
	The Project Supported by Natural Science Basic Research Plan in Shaanxi Province of China Program (No. 2025JC-YBQN-497) and Initiation Funds for High-level Talents Program of Xi’an International University (No. XAIU202518).
\end{acknowledgments}

\bibliography{ddr}

\end{document}